\documentclass[prl,twocolumn,superscriptaddress]{revtex4-1}

\usepackage{amsmath}
\usepackage{mathtools} % successor of amsmath
\usepackage{amssymb}
\usepackage{amsthm}
\usepackage{bm,upgreek} % Allows use of bold-italic greek letters
\usepackage{upgreek} % roma greek
\usepackage{xcolor}
\usepackage{graphicx}
\usepackage{epstopdf}
\usepackage[colorlinks,linkcolor=blue,anchorcolor=blue,citecolor=blue,urlcolor=blue]{hyperref}

\usepackage{multirow}
\usepackage{longtable,booktabs}

\newcommand{\nn}{{\nonumber}}
\newcommand{\bea}{\begin{eqnarray}}
\newcommand{\eea}{\end{eqnarray}}

\newcommand{\md}{\mathrm{d}}
\newcommand{\me}{\mathrm{e}}

\newcommand{\ru}{Sr$_2$RuO$_4${}}

\begin{document}

\title{Possible two-component spin-singlet pairings in Sr$_2$RuO$_4$}

\author{San-Jun Zhang}
\affiliation{National Laboratory of Solid State Microstructures $\&$ School of Physics, Nanjing University, Nanjing 210093, China}
\author{Da Wang} \email{dawang@nju.edu.cn}
\affiliation{National Laboratory of Solid State Microstructures $\&$ School of Physics, Nanjing University, Nanjing 210093, China}
\affiliation{Collaborative Innovation Center of Advanced Microstructures, Nanjing University, Nanjing 210093, China}
\author{Qiang-Hua Wang} \email{qhwang@nju.edu.cn}
\affiliation{National Laboratory of Solid State Microstructures $\&$ School of Physics, Nanjing University, Nanjing 210093, China}
\affiliation{Collaborative Innovation Center of Advanced Microstructures, Nanjing University, Nanjing 210093, China}

\begin{abstract}
Recent experiments suggest a multi-component pairing function in \ru, which appears to be inconsistent with the absence of an apparent cusp in the transition temperature ($T_c$) as a function of the uniaxial strain. 
We show, however, that the theoretical cusp in $T_c$ for a multi-component pairing can be easily smeared out by the spatial inhomogeneity of strain, and the experimental data can be reproduced qualitatively by a percolation model. This shed new light on multi-component pairings. We then perform a thorough group-theoretical classification of the pairing functions, taking the spin-orbit coupling into account. We list all $13$ types of two-component spin-singlet pairing functions, with $8$ of them belonging to the $E_g$ representation. In particular, we find two types of intra-orbital pairings in the $E_g$ representation  ($k_xk_z,k_yk_z$) are favorable in view of most existing experiments.
\end{abstract}
\maketitle

% experiments:
% NMR, muSR, Kerr, neutron, specific heat, thermoconductivity,
% uniaxial pressure, ultrasound, superfluid density, STM, disorder
% SQUID, half-quantized flux

%\section{Introduction}
\emph{Introduction}.
\ru{} is a layered perovskite superconductor isostructural to the cuprate La$_2$CuO$_4$, and was widely studied since its discovery \cite{born_1994}.
Muon spin rotation ($\mu$SR) \cite{muSR_1998,Grinenko_split_2021} and Kerr experiments \cite{Kerr_2006} indicate the time reversal symmetry is spontaneously broken in the superconducting state, suggesting that the pairing order parameter must have multiple components. This is indeed consistent with the ultrasound experiments \cite{ultrasound_2020,ultrasound2_2020}.
Theoretically, a symmetry-protected multi-component pairing function must belong to the two-fold degenerate $E_g$ or $E_u$ representation of the underlying $D_{4h}$ point group of \ru. 
The two types of pairing functions differ in parity. 
Early phase-sensitive probes \cite{Josephson_2004,Josephson_2006,half_fluxon_2011,Little-Parks_2017,magnetoresistivity_2020} suggest the pairing function transforms as $k_x+ik_y$ belonging to the $E_u$ representation, as was proposed in the early stage \cite{classification_1995,RMP_2003}.

However, more recent and refined experiments strongly challenge the $k_x+ik_y$ spin-triplet pairing. First, according to the Ginzburg-Landau (GL) theory, the two components of the order parameter in the $E_u$ (or even $E_g$) representation can couple to the uniaxial strain linearly, leading to a cusp-like feature in the superconducting transition temperature ($T_c$) as a function of the uniaxial strain. But no apparent cusp is observed experimentally \cite{pressure_2014,pressure_2017,pressure_2018,pressure_SQUID_2018}, and this appears to rule out the possibility of a multi-component order parameter. Second, if the pairing is spin-triplet, the Knight-shift should not drop below $T_c$ (at least for a magnetic field applied orthogonal to the so-called $d$-vector of the triplet). But it actually drops significantly in strained as well as unstrained samples in most recent and refined nuclear magnetic resonance (NMR) experiments \cite{NMR_2019,NMR_specific_heat_2020}. This, together with the polarized neutron experiment \cite{neutron_2020}, act strongly against spin-triplet pairing, hence, difficult to reconcile with the phase-sensitive experiments \cite{Leggett_symmetry_2020}.
Importantly, in the same type of samples the ultrasonic measurement reveals signature of a multi-component order parameter \cite{ultrasound_2020,ultrasound2_2020}. On one hand, this combines the behavior of Knight-shift to suggest spin-singlet pairing that transforms as the $E_g$ or even $E_u$ representation in the presence of spin-orbital coupling (SOC), but on the other hand seems to conflict with the absence of cusp in $T_c$ versus the uniaxial strain \cite{pressure_2014,pressure_2017,pressure_2018,pressure_SQUID_2018}. The current situation is therefore rather paradoxical, and motivates careful re-examination of the pairing symmetry in \ru{} \cite{classification_Sigrist_2019,classification_Senechal_2019,interorbital_An_2020,Yao_2018,Yao_2019,interorbital_LDA_2019,proposal_dg_2020}, regarding issues such as single- vs multi-component, spin-singlet vs -triplet, nodal vs nodeless quasiparticles, {\it etc}.
 
In this work, we try to reconcile the paradox and reinforce the possibility of the spin-singlet pairing in the degenerate $E_g$ representation. First, we realize that the strain in the reported sample is inhomogeneous, as seen in scanning superconducting quantum interference device (SQUID) experiment \cite{pressure_SQUID_2018}.
%and possibly also in the strong nematicity in mesoscopic samples above $T_c$ in recent Hall bar %experiment.\cite{Hall_2020}
We show by a percolation model that in the presence of an inhomogeneous background of strain, the cusp is absent or smeared out even if the pairing function is in the $E_g$ or $E_u$ representation. Combining the ultrasound, Knight-shift, neutron, $\mu$SR and Kerr experiments provides a consistent picture of multi-component pairing in the $E_g$ or $E_u$ representation. Since there remains various forms of $E_g$ and $E_u$ representations in the system with multiple orbitals and SOC, we then perform a thorough group-theoretical classification of the pairing functions in spin, orbital and momentum, and discuss the relevance of various pairing functions in view of the other experiments, such as superfluid density \cite{superfluid_density_2000}, specific heat \cite{specific_heat_2004,specific_heat_2018} and thermal transports \cite{universal_thermal_conductivity_2002,thermal_conductivity_2017}. We find two types of intra-orbital pairings in the $E_g$ representation  (transforming as $k_xk_z + i k_yk_z$) are most favorable in view of such experiments.

%\section{Strain effect}
\emph{Strain effect}.
We first discuss the impact of the recent experiments using strain as the tuning parameter. The uniaxial component of the strain, $\varepsilon = \epsilon_{xx}-\epsilon_{yy}$, where $\epsilon_{ab}$ is the element of the strain tensor, transforms as $B_{1g}$ under symmetry operations. Since \ru{} is $D_{4h}$ symmetric, the multi-component pairing order parameter must belong to the $E_g$ or $E_u$ representation, except for accidental degeneracy between one-dimensional representations (that will not interest us here). Let the two-component order parameter be $(\eta_1, \eta_2)$, which transforms as $(x,y)$ in the $E_u$ case, or $(xz,yz)$ in the $E_g$ case. In either case, it is clear that $|\eta_1|^2 -|\eta_2|^2$ also transforms as $B_{1g}$. Therefore, the order parameters can couple to $\varepsilon$ linearly as $\alpha\varepsilon(|\eta_1|^2-|\eta_2|^2)$, with a coefficient $\alpha$. Within the GL theory, this coupling lifts the degeneracy in the bare transition temperatures for the two components, and the modified transition temperature is the higher one, so that the change in $T_c$ behaves as $\delta T_c\propto|\varepsilon|$, resulting a cusp dependence in $\varepsilon$ \cite{pressure_2014,proposal_dg_2020}. Rather unexpectedly, no apparent cusp feature has been observed in strain experiments \cite{pressure_2014,pressure_2017,pressure_2018,pressure_SQUID_2018}.
%,pressure_heat_capacity_2020} 
This result appears to rule out the picture of multi-component order parameter in \ru, provided that the strain distribution is uniform in the sample. 

\begin{figure}
	\centering
	\includegraphics[width=0.45\textwidth]{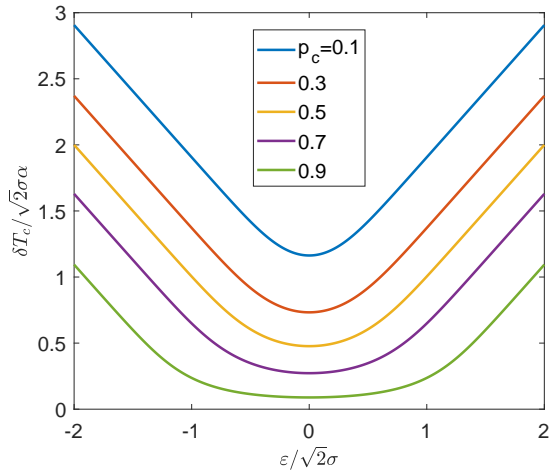}
	\caption{$\delta T_c$ vs uniaxial strain $\varepsilon$ at different percolation thresholds $p_c$. $\sigma$ is the standard deviation of the Gaussian distribution $\rho(\varepsilon_{\rm loc})$. }
	\label{fig:percolation}
\end{figure}

However, the recent scanning SQUID experiment \cite{pressure_SQUID_2018} shows that the local $T_c$, measured in different regions of the sample, reaches the minimum at different values of $\varepsilon$, although the minimal $T_c$ itself is only slightly changed. This fact clearly implies inhomogeneity of the strain distribution. To investigate the effect of such inhomogeneity, we assume a background strain $\varepsilon_{\rm loc}$, which distributes over the sample statistically with a probability density $\rho(\varepsilon_{\rm loc})$. 
The total strain at a specific spatial point is given by $\varepsilon_{eff}=\varepsilon+\varepsilon_{\rm loc}$, where $\varepsilon$ represents the applied (external) strain in experiments. This defines a local bare transition temperature $\tau(\varepsilon_{eff})=T_{c0}+|\alpha\varepsilon_{eff}|$, where $T_{c0}$ is the value of $T_c$ in the absence of any strain. We then have to deal with a system with $T_c$-inhomogeneity arising from the strain distribution. 
Notice that experimentally the position-dependent $T_c$ is determined by measuring the diamagnetic susceptibility within a ring of diameter $\sim2\mu m$ \cite{pressure_SQUID_2018}, which is much larger than the superconducting coherence length $\xi<100nm$ \cite{RMP_2003}.
Therefore, the superfluid induced diamagnetic signal can only be established if the associated large area has entered the superconducting state collectively. The large area justifies a statistical treatment of the strain distribution. By the simplest percolation model, we assume that superconductivity is achieved below $T_c$ if the statistical probability for $\tau(\varepsilon_{eff}) > T_c$ is above a percolation threashold $p_c$. In the classical percolation model, it is known that $p_c=1$ in one dimension, $p_c=0.5$ in two dimension, and is lower in higher dimensions. In our case it is reasonable to speculate that $0<p_c<0.5$, but its exact value is unimportant for qualitative purposes. 
In this picture, we can determine $\delta T_c = T_c-T_{c0}$ implicitly as
\begin{align}
\int\theta(\alpha|\varepsilon+\varepsilon_{\rm loc}|-\delta T_c)\rho(\varepsilon_{\rm loc})\md\varepsilon_{\rm loc}=p_c,
\label{eq:percolationTc}
\end{align}
where $\theta$ is the Heaviside step function. It turns out that the resulting $\delta T_c$ no longer develops cusp in the applied $\varepsilon$ as long as $p_c>0$. To see this point most straightforwardly, we can take derivative with respect to $\varepsilon$ in Eq.\ref{eq:percolationTc} to obtain
\begin{align}
\frac{\partial \delta T_c}{\partial \varepsilon}\left[ \rho\left(-\varepsilon+\frac{\delta T_c}{\alpha}\right) + \rho\left(-\varepsilon-\frac{\delta T_c}{\alpha}\right) \right] \nn\\
=\alpha \left[ \rho\left(-\varepsilon+\frac{\delta T_c}{\alpha}\right) - \rho\left(-\varepsilon-\frac{\delta T_c}{\alpha}\right) \right].
\end{align}
Clearly, $\partial \delta T_c/\partial \varepsilon=0$ at $\varepsilon=0$ as long as $\rho(\varepsilon_{\rm loc})$ is an even function. As a specific model, we assume a Gaussian distribution, $\rho(\varepsilon_{\rm loc})=\frac{1}{\sqrt{2\pi}\sigma}\me^{-{\varepsilon_{\rm loc}^2}/{2\sigma^2}}$. Then Eq.~\ref{eq:percolationTc} can be integrated out exactly, yielding
\begin{align}
\mathrm{erf}\left( \frac{\delta T_c-\alpha\varepsilon}{\sqrt{2}\sigma\alpha} \right) + \mathrm{erf}\left( \frac{\delta T_c+\alpha\varepsilon}{\sqrt{2}\sigma\alpha} \right) = 2(1-p_c),
\end{align}
where erf is the standard error function. In Fig.~\ref{fig:percolation}, we plot $\delta T_c$ vs $\varepsilon$ for various choices of $p_c$.
It can be seen that $\delta T_c$ depends smoothly on $\varepsilon$ unless the strain distribution width $\sigma$ goes to zero (uniform strain distribution).
Therefore, we have shown that {\it the smooth dependence of $T_c$ on small $\varepsilon$ cannot rule out the possibility of multi-component order parameter in \ru.}
Instead, combining the strain experiments with the $\mu$SR \cite{muSR_1998,Grinenko_split_2021}, neutron \cite{neutron_2020}, Kerr \cite{Kerr_2006} and ultrasound \cite{ultrasound_2020,ultrasound2_2020} experiments actually provides a consistent picture of spin-singlet pairing in the $E_g$ or $E_u$ representation.

We should remark that: (1) Experimentally, $T_c$ in different regions reaches minimum at different applied strains. This can be explained if the applied strain itself is nonuniform in the sample, such that the strain distribution is biased differently at different spatial regions (that are microscopically large but macroscopically small in the scale of coherence length); (2) The approximation of linear coupling to strain is valid only at small strains. Larger strains may modify the electronic structure significantly (because the Fermi level is close to the Van Hove singularity in the $\gamma$-band), and the effect inevitably goes beyond the linear approximation and beyond the scope of this work. 

\begin{widetext}
	\begin{center}
		\begin{table}
			\begin{tabular}{|c|c|c|c|}
				\hline \hline
				& 1&            $\lambda_{(23)}+\lambda_{(32)}$            &   $\lambda_{(13)}+\lambda_{(31)}$     \\ \cline{2-4}
				& 2&  $k_xk_y(k_x^2-k_y^2)(\lambda_{(13)}+\lambda_{(31)})$   &   $-k_xk_y(k_x^2-k_y^2)(\lambda_{(23)}+\lambda_{(32)})$   \\ \cline{2-4}
				&3 &     $(k_x^2-k_y^2)(\lambda_{(23)}+\lambda_{(32)})$ & $-(k_x^2-k_y^2)(\lambda_{(13)}+\lambda_{(31)})$     \\ \cline{2-4}
				$E_g$  &4 &         $k_xk_y(\lambda_{(13)}+\lambda_{(31)})$         &   $k_xk_y(\lambda_{(23)}+\lambda_{(32)})$   \\ \cline{2-4}
				&5 &                 $k_xk_z(\lambda_{(11)}+\lambda_{(22)})$                  &   $k_yk_z(\lambda_{(11)}+\lambda_{(22)})$   \\ \cline{2-4}
				&6 &                 $k_yk_z(\lambda_{(12)}+\lambda_{(21)})$                  &  $k_xk_z(\lambda_{(12)}+\lambda_{(21)})$   \\ \cline{2-4}
				&7 &                         $k_xk_z\lambda_{(33)}$                         &  $k_yk_z\lambda_{(33)}$   \\ \cline{2-4}
				&8 &                 $k_xk_z(\lambda_{(11)}-\lambda_{(22)})$                  &   $-k_yk_z(\lambda_{(11)}-\lambda_{(22)})$    \\ \hline
				&9 & $k_xk_yk_z(k_x^2-k_y^2)(\lambda_{(13)}-\lambda_{(31)})$ &  $k_xk_yk_z(k_x^2-k_y^2)(\lambda_{(23)}-\lambda_{(32)})$ \\ \cline{2-4}
				&10&          $k_z(\lambda_{(23)}-\lambda_{(32)})$           &  $k_z(\lambda_{(13)}-\lambda_{(31)})$   \\ \cline{2-4}
				$E_u$  &11&       $k_xk_yk_z(\lambda_{(13)}-\lambda_{(31)})$        & $k_xk_yk_z(\lambda_{(13)}-\lambda_{(31)})$  \\ \cline{2-4}
				&12&    $k_z(k_x^2-k_y^2)(\lambda_{(23)}-\lambda_{(32)})$    & $k_z(k_x^2-k_y^2)(\lambda_{(13)}-\lambda_{(31)})$  \\ \cline{2-4}
				&13&                    $k_y(\lambda_{(12)}-\lambda_{(21)})$                     &  $k_x(\lambda_{(12)}-\lambda_{(21)})$   \\ \hline
			\end{tabular}
			\caption{Spin-singlet pairing functions in the $E_g$ and $E_u$ representations. Here $\lambda_{(ab)}$ denotes a matrix in the orbital basis, with the elements $\lambda^{ij}_{(ab)}=\delta_{ia}\delta_{jb}$. On each row, the two basis functions transform as $(xz,yz)$, respectively, in the $E_g$ representation, and as $(x,y)$ in the $E_u$ representation. The identity matrix $\sigma_0$ in the spin basis, for spin-singlet pairing, is omitted for brevity. 
				\label{table:singlet}}
		\end{table}
	\end{center}
\end{widetext}

%\section{group classification}
\emph{Group classification}.
\ru{} is a multi-orbital system and its low energy bands are dominantly described by three $t_{2g}$ orbitals ($d_{xz}$, $d_{yz}$, $d_{xy}$) \cite{RMP_2003}.
On the other hand, ARPES has shown its band structures are strongly affected by the SOC effect \cite{ARPES_2014}.
Therefore, a thorough group classification of the possible pairing functions is needed, taking into account the $t_{2g}$ orbitals, the atomic SOC, and the $D_{4h}$ group. A general pairing Hamiltonian for electrons at momenta $\0k$ and $-\0k$ can be written as $\psi_\0k^\dagger \Delta(\0k) i\sigma_2\psi_{-\0k}^{\dagger,t} + {\rm h.c.}$, where $\psi_\0k$ is a multi-component spinor composed of electron annihilation operators of all internal degrees of freedom, such as orbital and spin, $\Delta(\0k)$ is a matrix function in the orbital and spin bases, and $\sigma_{\mu=0,1,2,3}$ will henceforth denote the identity and Pauli matrices acting on spins. Under a group operation $g\in D_{4h}$, the spinor transforms as $\psi_\0k'=U_g\psi_{g^{-1}\0k}$, where $U_g$ encodes the $\0k$-independent transformation on orbitals and spins. Correspondingly, the pairing matrix transforms as $\Delta_\0k' = U_g \Delta_{g^{-1}\0k} U_g^\dagger$, where we used the fact that $U_g = T U_g T^{-1}$ for transformation on spins and on real orbital bases, with $T=i\sigma_2 \+K$ the time-reversal operator (and $\+K$ the complex conjugation operator). 
The pairing matrix can be written as a linear superposition of the tensor products $\Lambda_o\otimes\Lambda_s\otimes f_\0k$, where $\Lambda_{o (s)}$ is a matrix describing a bilinear in the orbital (spin) basis, and $f_\0k$ is a function of $\0k$. 
To set up notations, we define $\lambda_{(ab)}$ as a self-explaining matrix in the orbital basis $(d_{xz}, d_{yz}, d_{xy})$ such that its elements read $\lambda_{(ab)}^{ij}=\delta_{ia}\delta_{jb}$. Such matrices can be used to expand $\Lambda_o$ and form irreducible representations. For example, $\lambda_{(11)}+\lambda_{(22)}\sim A_{1g}$, $\lambda_{(11)} -\lambda_{(22)}\sim B_{1g}$, etc. The matrix $\Lambda_s$ is expanded by $\sigma_\mu$, with $\sigma_0$ ($\sigma_{1,2,3}$) representing the spin-singlet (triplet) component(s). Under $D_{4h}$ these matrices transform as $\sigma_0\sim A_{1g}$, $\sigma_3\sim A_{2g}$, $(\sigma_1, \sigma_2)\sim E_{g}$. (Note that under time-reversal, $\sigma_0$ is invariant while $\sigma_{1,2,3}$ changes sign.)  
The classification of the function $f_\0k$ is standard. The details of the separate classifications are presented in the appendix. Finally, the entire pairing function is classified by decomposing the tensor products of the separate irreducible representations. The complete results are also presented in the appendix. Note that it is possible that multiple pairing functions with either spin-singlet or -triplet, transform identically as the same irreducible reprentation. Such pairing functions could mix, but this does not mean they have to, since the extent of mixing of such pairing functions is not dictated by symmetry alone. 

Here we focus on spin-singlet $E_g$ and $E_u$ representations, along the line of the previous discussions of the experiments. Such pairing functions are listed in Table \ref{table:singlet}. Note that if necessary, each pairing function could be multiplied by an additional $A_{1g}$ factor function of $\0k$ to describe pairing on longer bonds. There are $8$ pairing functions in the $E_g$ representation, and $5$ in the $E_u$ representation.
(Note that $E_u$ spin-singlet is allowed if the pairing function is odd under orbital exchange.)
Among these pairing functions, only three of them (No. 5, 7, and 8) in the $E_g$ representation are intra-orbital pairing, while all the others involve inter-orbital pairing. If the pairing arises from electron-electron correlations, the orbital-wise matrix elelment effect in the overlap between Bloch states should render inter-orbital pairing less relevant. In this case, we may speculate that the above three $E_g$ pairing functions are the most important. Once the dominant pairing functions in the $E_g$ representation are realized, the others in the same representation may be induced by subleading correlation effects, so for sufficient generality, we include all of the $E_g$ functions in the list. In this setting, we can write the two degenerate general pairing functions in the $E_g$ representation explicitly as, in the orbital basis, 

\begin{widetext}	
\begin{align}
\Delta_{xz}(\0k)&=\begin{bmatrix}
(d_5+d_8)k_xk_z & d_6k_yk_z & d_4k_xk_y+d_2k_xk_y(k_x^2-k_y^2) \\
d_6k_yk_z& (d_5-d_8)k_xk_z & d_3(k_x^2-k_y^2)+d_1 \\
d_4k_xk_y+d_2k_xk_y(k_x^2-k_y^2)&d_3(k_x^2-k_y^2) +d_1& d_7k_xk_z
\end{bmatrix}, \nn\\
\Delta_{yz}(\0k)&=\begin{bmatrix}
(d_5-d_8)k_yk_z & d_6k_xk_z & -d_3(k_x^2-k_y^2)+d_1 \\
d_6k_xk_z& (d_5+d_8)k_yk_z & d_4k_xk_y-d_2k_xk_y(k_x^2-k_y^2) \\
-d_3(k_x^2-k_y^2)+d_1&d_4k_xk_y-d_2k_xk_y(k_x^2-k_y^2)& d_7k_yk_z
\end{bmatrix}.
\end{align}
\end{widetext}
Here $d_{1\sim 8}$ are coefficients multiplying the respective $E_g$ functions in Table~\ref{table:singlet}. The relative ratios among these coefficients depend on the microscopic details. (Once the relative ratio is fixed, in the GL theory, the global coefficients act as order parameters, and carry, or transform as, the same $E_g$ representation.) Below the transition temperature, it is usually favorable for the two degenerate pairing functions, $\Delta_{xz}(\0k)$ and $\Delta_{yz}(\0k)$ to combine into the time reversal symmetry breaking form, $\Delta(\0k)=\Delta_{xz}(\0k)\pm i\Delta_{yz}(\0k)$, to maximize the pairing gaps on the Fermi surface and gain energy.

\begin{figure}
\includegraphics[width=0.48\textwidth]{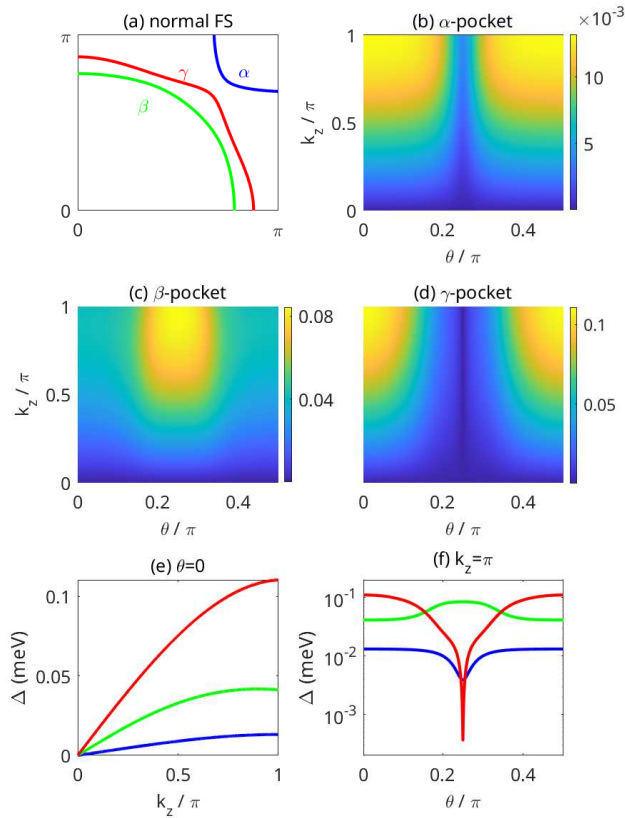}
\caption{\label{fig:gap} A possible pairing of $E_g$ with $d_7=0.01$ and $d_8=0.005$ (in unit of mRy). The normal state FS at $k_z=0$ is plotted in (a) to define the $\alpha$, $\beta$, $\gamma$ pockets. In (b) to (d), the quasiparticle gaps are plotted on the $(\theta,k_z)$ plane for each pocket. $\theta$ is defined as the azimuth angle relative to the center of each pocket. In special, the $k_z$-dependence at $\theta=0$ is explicitly shown in (e) and $\theta$-dependence at $k_z=\pi$ in (f).
}
\end{figure}

\emph{Gap structure}.
We now discuss the quasiparticle excitations subject to the above spin-singlet $E_g$ pairing functions. The Bogoliubov-de Gennes Hamiltonian in the Nambu basis $\Psi_\0k^\dagger = (\psi_{\0k}^\dagger,-\psi_{-\0k}^t i\sigma_2)$ is, 
\begin{align}
H=\sum_{\0k}\Psi_\0k^\dagger \begin{bmatrix}
h_\0k & \Delta(\0k) \\ \Delta^\dagger(\0k) & -Th_{-\0k}T^{-1}
\end{bmatrix} \Psi_\0k,
\end{align}
where $h_\0k$ is the normal state single-particle Hamiltonian taken from Ref.~\cite{Mazin_2006,ARPES_2014} with atomic SOC, and $\Delta(\0k)=\Delta_{xz}(\0k)+i\Delta_{yz}(\0k)$ (tensor producted implicitly by $\sigma_0$).
In order to obtain the desired $(k_xk_z,k_yk_z)$ intra-orbital pairing as discussed above, we consider the nearest-neighbour inter-layer pairings on bonds $(\pm a/2,\pm a/2,\pm c/2)$ where $a$ and $c$ are in-plane and out-of-plane lattice constants.
The simple form factor $f_\0k$ is replaced by the corresponding lattice harmonics, e.g., $k_x k_z\rightarrow \sin (k_x a/2) \sin(k_z c/2)$, $k_x^2-k_y^2\rightarrow \cos(k_x a/2)-\cos(k_y a/2)$, etc. 
The quasiparticle gaps for different pairings, characterized by the coefficients $d_{i=1,2,\cdots,8}$, can be found in the appendix.
For all the $E_g$ pairings, there is a horizontal nodal line at $k_z=0$ by symmetry. This is consistent with the specific heat \cite{specific_heat_2018} and neutron \cite{neutron_horizontal_2020} experiments.
When inter-orbital pairing is included, the horizontal nodal line may expand into a nodal torus, forming the Bogoliubov Fermi surface \cite{Bogoliubov_FS_2017} as shown in the appendix. This would generate a finite zero-energy quasiparticle density of states, which is however inconsistent with the universal thermal conductivity \cite{universal_thermal_conductivity_2002,thermal_conductivity_2017}, that can only arise if the energy gap is nodal or quasi-nodal \cite{Lee_universal_2000,QHW_universal_2018,QHW_FRG_2019}.
Furthermore, the substantial nonzero c-axis thermal conductivity in the $T=0$K limit \cite{thermal_conductivity_2017} was taken as a strong evidence to rule out the ``simple'' $k_z=0$ horizontal nodal line picture.
Recently, the STM experiment \cite{STM_2020} also indicates the existence of a vertical nodal line (or gap minima) along (11)-direction.
{\it Taking these together, we find $(d_7,d_8)$-pairing may be the most relevant.}
%Its gap structure is shown in Fig.~\ref{fig:gap}.
With parametrization $d_7=2d_8=0.01$mRy, the quasiparticle gaps on three pockets as defined in Fig.~\ref{fig:gap}(a) are plotted on $(\theta,k_z)$ plane in Figs.~\ref{fig:gap}(b) to \ref{fig:gap}(d), respectively, where $\theta$ is the azimuthal angle relative to the pocket center. In particular, for clarity, the gaps along two cuts $\theta=0$ and $k_z=\pi$ are given in Figs.~\ref{fig:gap}(e) and \ref{fig:gap}(f).
Clearly, in addition to the horizontal nodal lines for all three pockets, there is a very deep gap minima at $\theta=\pi/4$ on the $\gamma$-pocket.
This quasi-node stems from the effect of SOC, which causes the Bloch state at the Fermi angle $\theta=\pi/4$ on the $\gamma$-pocket to be dominated by the $d_{xz}$ and $d_{yz}$ components, while the pairing of the latter two orbitals ($\lambda_{(11)}-\lambda_{(22)}$ contributing a $B_{1g}$ factor within the $E_g$ representation) has an exact node at $\theta=\pi/4$. This quasi-node naturally explains the observed universal thermal conductivity \cite{QHW_universal_2018,QHW_FRG_2019,Dodaro_gapminima_2018}.

Another issue raised by the experiment \cite{thermal_conductivity_2017} is the superficial mutual scaling of the in-plane and out-of-plane thermal conductivities, shown as normalized $\kappa_{ab}/T$ and $\kappa_c/T$ versus the magnetic field. This was taken as the basis to exclude the horizontal nodal line, since the Fermi velocity here is in-plane hence  contributes to $\kappa_{ab}/T$ but not $\kappa_c/T$. While it is reasonable in the presence of horizontal node alone, the argument needs to be re-examined if the vertical node (or quasi-node) is also present. In the latter case, both in-plane and out-of-plane thermal transports are possible, and both types of nodes (quasi-nodes) are subject to the Volovik effect \cite{Volovik_DOS_1993} which induces a zero-energy density of states (DOS) $\rho(0)=\+N_B \propto \sqrt{B}$, where $B$ is the magnetic field. In the presence of impurity scattering rate $\gamma$, the effective DOS is given by $\rho_{\rm eff}(0)\sim\text{max}(\gamma,\+N_B)$. Therefore, when $\+N_B >\gamma$, both $\kappa_{ab}/T$ and $\kappa_c/T$ are proportional to $\+N_B$, which explains the observed mutual scaling.

%\section{Summary}
\emph{Summary}.
In this work, we first resolve the paradox between the multi-component pairings and the uniaxial strain experiments.
Then by performing a thorough group classification based on the D$_{4h}$ group with SOC included and by carefully examining different experiments, we conclude that the $E_g$-pairing is the most probable symmetry for \ru, namely $(d_{xz},d_{yz})$-wave [transforming as $(k_xk_z,k_yk_z)$].
In particular, we point out the spin-singlet intra-orbital pairings dominated by $(k_xk_z,k_yk_z)\lambda_{(33)}$ and $(k_xk_z,k_yk_z)(\lambda_{(11)}-\lambda_{(22)})$ are compatible to most known experiments.

It is important to ask what is the pairing mechanism that would cause the inter-layer $E_g$-pairing, which would possibly also explain why $T_c$ of \ru{} is much lower than cuprates.
In this regard, a careful study of the three dimensional three-orbital Hubbard model with SOC may shed new light on the underlying pairing mechanism \cite{Roising_hubbard_2019,interorbital_kSOC_2020}.
Another remaining question is how to explain the existing phase sensitive experiments \cite{Josephson_2004,Josephson_2006,half_fluxon_2011,magnetoresistivity_2020,Little-Parks_2017}  and reconcile the singlet nature of the pairing seen in the NMR experiments. This deserves further studies both theoretically and experimentally.

\emph{Acknowledge}.
This work is supported by National Natural Science Foundation of China (under Grant Nos. 11874205 and 11574134) and National Key Research and Development Program of China (under Grant No. 2016YFA0300401).

%\nocite{*}
\bibliography{Sr2RuO4}

\clearpage
\appendix
%\begin{widetext}
\onecolumngrid
\section{Appendix}
In this appendix, we provide the results of a thorough group classification of \ru{} based on the D$_{4h}$ group with SOC included. Then, the gap structures of typical spin-singlet $E_g$ pairings are given.

\section{Group classification}

Following the notations in the main text, the pairing matrix can be written as a tensor product $\Lambda_o\otimes\Lambda_s\otimes f_\0k$, where $\Lambda_o$, $\Lambda_s$ and $f_\0k$ are for orbital, spin and momentum, respectively. $\Lambda_o$ can be expanded on $\lambda_{(ab)}$ and $\Lambda_s$ on $\sigma_\mu$, where $\lambda_{(ab)}$ denotes the matrix with $(ij)$-element given by $\lambda_{(ab)}^{ij}=\delta_{ia}\delta_{jb}$ and $\sigma_{0}$($\sigma_{1,2,3}$) represent(s) spin singlet (triplet).
All of the three parts transform as independent irreducible representations as listed in table \ref{table:IRPfunctions}. After obtaining these representations, we apply group product to obtain a thorough list of all 148 pairings as listed in table \ref{table:full}.

\begin{table}[h]
\begin{center}
\caption{Irreducible representations of the pairing matrix in orbital, spin and momentum spaces, respectively. The total pairing can be any tensor product of these three parts with odd parity in total.
\label{table:IRPfunctions}
}
\begin{tabular}{|c|c|c|c|}
	\hline
	         & {orbital} &           {spin}           &          $f(k)$          \\ \hline
	$A_{1g}$ &   $\begin{array}{c}\lambda_{(11)}+\lambda_{(22)} \\ \lambda_{(33)} \end{array}$    &      $\sigma_0$       &            $1$            \\ \hline
	$A_{2g}$ &    $\lambda_{(12)}-\lambda_{(21)}$    &      $\sigma_3$       &  $k_xk_y(k_x^2-k_y^2)$   \\ \hline
	$B_{1g}$ &   $\lambda_{(11)}-\lambda_{(22)}$    &            $-$             &      $k_x^2-k_y^2$       \\ \hline
	$B_{2g}$ &   $\lambda_{(12)}+\lambda_{(21)}$   &            $-$             &         $k_xk_y$         \\ \hline
	$E_{g}$  &   $\begin{array}{c} (\lambda_{(13)}+\lambda_{(31)},\lambda_{(23)}+\lambda_{(32)}) \\ (\lambda_{(13)}-\lambda_{(31)},\lambda_{(23)}-\lambda_{(32)}) \end{array}$    & $(\sigma_1,\sigma_2)$ &    $(k_xk_z,k_yk_z)$     \\ \hline
	$A_{1u}$ &   $-$   &            $-$             & $k_xk_yk_z(k_x^2-k_y^2)$ \\ \hline
	$A_{2u}$ &   $-$   &            $-$             &          $k_z$           \\ \hline
	$B_{1u}$ &   $-$   &            $-$             &       $k_xk_yk_z$        \\ \hline
	$B_{2u}$ &   $-$   &            $-$             &    $k_z(k_x^2-k_y^2)$    \\ \hline
	$E_{u}$  &   $-$   &            $-$             &       $(k_x,k_y)$        \\ \hline
\end{tabular}
\end{center}
\end{table}

%-------------------------------------------------
%\renewcommand\arraystretch{1.}
%\setlength\LTleft{0in}
%\setlength\LTright{0in}
\begingroup
\nopagebreak
\LTcapwidth=\textwidth
\squeezetable
\begin{longtable}{|c|ccc|}
\caption{A full classification of \ru. 
%The pairings are represented by a generalized $\0d$-vector: $d_{\mu,ij}(k)=\av{\psi_{i\alpha}(-k)(\sigma_\mu i\sigma_2)_{\alpha\beta}\lambda_{ij}\psi_{j\beta}(k)}$. The index $i,j=1,2,3$ correspond to $d_{xz},d_{yz},d_{xy}$ orbitals. 
In general, a pairing can written as a linear superposition of the tensor products $\lambda_{(ab)}\otimes\sigma_\mu\otimes f_\0k$.
Notice that $f_\0k$ only give the lowest order lattice harmonics. In this table, for compactness, we use the notation $\lambda_{ab}$ to represent the matrix $\lambda_{(ab)}$ as defined in the main text. \label{table:full}
}\\ 
\hline
irrep.	& $\sigma_0$ & $\sigma_3$ & $(\sigma_1,\sigma_2)$ \\
\hline
%\endfirsthead
%
%\hline
%irrep.	& $\sigma_0 \cdot i\sigma_2$ & $\sigma_3 \cdot i\sigma_2$ & $(\sigma_1,\sigma_2)\cdot i\sigma_2$ \\
%\hline
\endhead

					\multirow{5}{*}{$A_{1g}$} & $\lambda_{11}+\lambda_{22} $ & \multirow{5}{*}{\shortstack{$(\lambda_{12}-\lambda_{21})\sigma_3$ \\ $[k_{x}k_{z}(\lambda_{13}-\lambda_{31})-k_{y}k_{z}(\lambda_{23}-\lambda_{32})]\sigma_3$}} & $(\lambda_{12}-\lambda_{21})(k_{x}k_{z}\sigma_1-k_{y}k_{z}\sigma_2)$ \\
					& $\lambda_{33} $  && $(\lambda_{13}-\lambda_{31})\sigma_1+(\lambda_{23}-\lambda_{32})\sigma_2$  \\
					& $(k_{x}^{2}-k_{y}^{2})(\lambda_{11}-\lambda_{22}) $ && $k_{x}k_{y}(k_{x}^{2}-k_{y}^{2})[(\lambda_{13}-\lambda_{31})\sigma_2-(\lambda_{23}-\lambda_{32})\sigma_1]$ \\
					& $k_{x}k_{y}(\lambda_{12}+\lambda_{21}) $ && $(k_{x}^{2}-k_{y}^{2})[(\lambda_{13}-\lambda_{31})\sigma_1-(\lambda_{23}-\lambda_{32})\sigma_2]$ \\
					& $k_{y}k_{z}(\lambda_{13}+\lambda_{31})+k_{x}k_{z}(\lambda_{23}+\lambda_{32}) $ && $k_{x}k_{y}[(\lambda_{13}-\lambda_{31})\sigma_2+(\lambda_{23}-\lambda_{32})\sigma_1]$ \\  \hline
					\multirow{5}{*}{$A_{2g}$} &  $k_{x}k_{y}(k_{x}^{2}-k_{y}^{2})(\lambda_{11}+\lambda_{22}) $ & \multirow{5}{*}{\shortstack{$k_{x}k_{y}(k_{x}^{2}-k_{y}^{2})(\lambda_{12}-\lambda_{21})\sigma_3$ \\
							$[k_{y}k_{z}(\lambda_{13}-\lambda_{31})+k_{x}k_{z}(\lambda_{23}-\lambda_{32})]\sigma_3$}} & $(\lambda_{12}-\lambda_{21})(k_{y}k_{z}\sigma_1+k_{x}k_{z}\sigma_2)$ \\
					& $k_{x}k_{y}(k_{x}^{2}-k_{y}^{2})\lambda_{33} $ && $(\lambda_{13}-\lambda_{31})\sigma_2-(\lambda_{23}-\lambda_{32})\sigma_1$ \\
					& $k_{x}k_{y}(\lambda_{11}-\lambda_{22}) $ && $k_{x}k_{y}(k_{x}^{2}-k_{y}^{2})[(\lambda_{13}-\lambda_{31})\sigma_1+(\lambda_{23}-\lambda_{32})\sigma_2]$ \\
					& $(k_{x}^{2}-k_{y}^{2})(\lambda_{12}+\lambda_{21}) $&& $(k_{x}^{2}-k_{y}^{2})[(\lambda_{13}-\lambda_{31})\sigma_2+(\lambda_{23}-\lambda_{32})\sigma_1]$ \\
					& $k_{x}k_{z}(\lambda_{13}+\lambda_{31})-k_{y}k_{z}(\lambda_{23}+\lambda_{32}) $&& $k_{x}k_{y}[(\lambda_{13}-\lambda_{31})\sigma_1-(\lambda_{23}-\lambda_{32})\sigma_2]$ \\  \hline
					\multirow{5}{*}{$B_{1g}$} & $(k_{x}^{2}-k_{y}^{2})(\lambda_{11}+\lambda_{22}) $ & \multirow{5}{*}{\shortstack{$(k_{x}^{2}-k_{y}^{2})(\lambda_{12}-\lambda_{21})\sigma_3$  \\
							$[k_{x}k_{z}(\lambda_{13}-\lambda_{31})+k_{y}k_{z}(\lambda_{23}-\lambda_{32})]\sigma_3$}}& $k_{x}k_{z}(\lambda_{12}-\lambda_{21})\sigma_1+k_{y}k_{z}(\lambda_{12}-\lambda_{21})\sigma_2$ \\
					& $(k_{x}^{2}-k_{y}^{2})\lambda_{33} $ && $(\lambda_{13}-\lambda_{31})\sigma_1-(\lambda_{23}-\lambda_{32})\sigma_2$ \\
					& $\lambda_{11}-\lambda_{22} $ && $k_{x}k_{y}(k_{x}^{2}-k_{y}^{2})[(\lambda_{13}-\lambda_{31})\sigma_2+(\lambda_{23}-\lambda_{32})\sigma_1]$ \\
					& $k_{x}k_{y}(k_{x}^{2}-k_{y}^{2})(\lambda_{12}+\lambda_{21}) $ && $(k_{x}^{2}-k_{y}^{2})[(\lambda_{13}-\lambda_{31})\sigma_1+(\lambda_{23}-\lambda_{32})\sigma_2]$\\
					& $k_{y}k_{z}(\lambda_{13}+\lambda_{31})-k_{x}k_{z}(\lambda_{23}+\lambda_{32})$ && $k_{x}k_{y}[(\lambda_{13}-\lambda_{31})\sigma_2-(\lambda_{23}-\lambda_{32})\sigma_1]$ \\  \hline
					\multirow{5}{*}{$B_{2g}$} & $k_{x}k_{y}(\lambda_{11}+\lambda_{22}) $ & \multirow{5}{*}{\shortstack{$k_{x}k_{y}(\lambda_{12}-\lambda_{21})\sigma_3$  \\
							$[k_{y}k_{z}(\lambda_{13}-\lambda_{31})-k_{x}k_{z}(\lambda_{23}-\lambda_{32})]\sigma_3$}}& $(\lambda_{12}-\lambda_{21})(k_{y}k_{z}\sigma_1-k_{x}k_{z}\sigma_2)$ \\
					& $k_{x}k_{y}\lambda_{33} $ && $(\lambda_{13}-\lambda_{31})\sigma_2+(\lambda_{23}-\lambda_{32})\sigma_1$ \\
					& $k_{x}k_{y}(k_{x}^{2}-k_{y}^{2})(\lambda_{11}-\lambda_{22}) $ && $k_{x}k_{y}(k_{x}^{2}-k_{y}^{2})[(\lambda_{13}-\lambda_{31})\sigma_1-(\lambda_{23}-\lambda_{32})\sigma_2]$ \\
					& $\lambda_{12}+\lambda_{21} $ && $(k_{x}^{2}-k_{y}^{2})[(\lambda_{13}-\lambda_{31})\sigma_2-(\lambda_{23}-\lambda_{32})\sigma_1]$ \\
					& $k_{x}k_{z}(\lambda_{13}+\lambda_{31})-k_{y}k_{z}(\lambda_{23}+\lambda_{32}) $ && $k_{x}k_{y}[(\lambda_{13}-\lambda_{31})\sigma_1+(\lambda_{23}-\lambda_{32})\sigma_2]$ \\  \hline
					\multirow{8}{*}{$E_{g}$} &  $k_{z}(k_{x},k_{y})(\lambda_{11}+\lambda_{22}) $ & \multirow{8}{*}{\shortstack{$k_{z}(k_{x},k_{y})(\lambda_{12}-\lambda_{21})\sigma_3$  \\
							$[\lambda_{13}-\lambda_{31},-(\lambda_{23}-\lambda_{32})]\sigma_3$ \\$k_{x}k_{y}(k_{x}^{2}-k_{y}^{2})(\lambda_{23}-\lambda_{32},\lambda_{13}-\lambda_{31})\sigma_3$ \\$(k_{x}^{2}-k_{y}^{2})(\lambda_{13}-\lambda_{31},\lambda_{23}-\lambda_{32})\sigma_3$ \\$k_{x}k_{y}[\lambda_{23}-\lambda_{32},-(\lambda_{13}-\lambda_{31})]\sigma_3$ }} & $(\lambda_{12}-\lambda_{21})(\sigma_1,-\sigma_2)$  \\
					& $k_{z}(k_{x},k_{y})\lambda_{33} $ && $k_{x}k_{y}(k_{x}^{2}-k_{y}^{2})(\lambda_{12}-\lambda_{21})(\sigma_2,\sigma_1)$ \\
					& $k_{z}(k_{x},-k_{y})(\lambda_{11}-\lambda_{22}) $ && $(k_{x}^{2}-k_{y}^{2})(\lambda_{12}-\lambda_{21})(\sigma_1,\sigma_2)$ \\
					& $k_{z}(k_{y},k_{x})(\lambda_{12}+\lambda_{21}) $ && $k_{x}k_{y}(\lambda_{12}-\lambda_{21})(\sigma_2,-\sigma_1)$ \\
					& $(\lambda_{23}+\lambda_{32},\lambda_{13}+\lambda_{31}) $ && $k_{z}(k_{x},k_{y})[(\lambda_{13}-\lambda_{31})\sigma_1+(\lambda_{23}-\lambda_{32})\sigma_2]$ \\
					& $k_{x}k_{y}(k_{x}^{2}-k_{y}^{2})[\lambda_{13}+\lambda_{31},-(\lambda_{23}+\lambda_{32})] $ && $k_{z}(k_{y},-k_{x})[(\lambda_{13}-\lambda_{31})\sigma_2-(\lambda_{23}-\lambda_{32})\sigma_1]$ \\
					& $(k_{x}^{2}-k_{y}^{2})[\lambda_{23}+\lambda_{32},-(\lambda_{23}+\lambda_{32})] $ && $k_{z}(k_{x},-k_{y})[(\lambda_{13}-\lambda_{31})\sigma_1-(\lambda_{23}-\lambda_{32})\sigma_2]$ \\
					& $k_{x}k_{y}(\lambda_{13}+\lambda_{31},\lambda_{23}+\lambda_{32}) $ && $k_{z}(k_{y},k_{x})[(\lambda_{13}-\lambda_{31})\sigma_2+(\lambda_{23}-\lambda_{32})\sigma_1]$ \\  	\hline
					\multirow{8}{*}{$A_{1u}$}	& \multirow{8}{*}{\shortstack{$k_{z}(\lambda_{12}-\lambda_{21}) $  \\
							$k_{x}(\lambda_{13}-\lambda_{31})-k_{y}(\lambda_{23}-\lambda_{32}) $}} & \multirow{8}{*}{\shortstack{$k_{z}(\lambda_{11}+\lambda_{22})\sigma_3$  \\
							$k_{z}\lambda_{33}\sigma_3$  \\
							$k_{z}(k_{x}^{2}-k_{y}^{2})(\lambda_{11}-\lambda_{22})\sigma_3$ \\
							$k_{x}k_{y}k_{z}(\lambda_{12}+\lambda_{21})\sigma_3$ \\
							$[k_{y}(\lambda_{13}+\lambda_{31})+k_{x}(\lambda_{23}+\lambda_{32})]\sigma_3$}} & $(\lambda_{11}+\lambda_{22})(k_{x}\sigma_1-k_{y}\sigma_2)$ \\
					&&& $\lambda_{33}(k_{x}\sigma_1-k_{y}\sigma_2)$ \\
					&&& $(\lambda_{11}-\lambda_{22})(k_{x}\sigma_1+k_{y}\sigma_2)$ \\
					&&& $(\lambda_{12}+\lambda_{21})(k_{y}\sigma_1-k_{x}\sigma_2)$\\
					&&& $k_{x}k_{y}k_{z}(k_{x}^{2}-k_{y}^{2})[(\lambda_{13}+\lambda_{31})\sigma_1+(\lambda_{23}+\lambda_{32})\sigma_2]$\\
					&&& $k_{z}[(\lambda_{13}+\lambda_{31})\sigma_2-(\lambda_{23}+\lambda_{32})\sigma_1]$\\
					&&& $k_{x}k_{y}k_{z}[(\lambda_{13}+\lambda_{31})\sigma_1-(\lambda_{23}+\lambda_{32})\sigma_2]$\\
					&&& $k_{z}(k_{x}^{2}-k_{y}^{2})[(\lambda_{13}+\lambda_{31})\sigma_2+(\lambda_{23}+\lambda_{32})\sigma_1]$\\  \hline
					\multirow{8}{*}{$A_{2u}$}	& \multirow{8}{*}{\shortstack{$k_{x}k_{y}k_{z}(k_{x}^{2}-k_{y}^{2})(\lambda_{12}-\lambda_{21}) $ \\
							$k_{y}(\lambda_{13}-\lambda_{31})+k_{x}(\lambda_{23}-\lambda_{32})$}} & \multirow{8}{*}{\shortstack{$k_{x}k_{y}k_{z}(k_{x}^{2}-k_{y}^{2})(\lambda_{11}+\lambda_{22})\sigma_3$ \\
							$k_{x}k_{y}k_{z}(k_{x}^{2}-k_{y}^{2})\lambda_{33}\sigma_3$ \\
							$k_{x}k_{y}k_{z}(\lambda_{11}-\lambda_{22})\sigma_3$    \\
							$k_{z}(k_{x}^{2}-k_{y}^{2})(\lambda_{12}+\lambda_{21})\sigma_3$  \\
							$[k_{x}(\lambda_{13}+\lambda_{31})-k_{y}(\lambda_{23}+\lambda_{32})]\sigma_3$}} & $(\lambda_{11}+\lambda_{22})(k_{y}\sigma_1+k_{x}\sigma_2)$ \\
					&&& $k_{y}\lambda_{33}\sigma_1+k_{x}\lambda_{33}\sigma_2$ \\
					&&& $(\lambda_{11}-\lambda_{22})(k_{y}\sigma_1-k_{x}\sigma_2)$ \\
					&&& $(\lambda_{12}+\lambda_{21})(k_{x}\sigma_1+k_{x}\sigma_2)$\\
					&&& $k_{x}k_{y}k_{z}(k_{x}^{2}-k_{y}^{2})[(\lambda_{13}+\lambda_{31})\sigma_2-(\lambda_{23}+\lambda_{32})\sigma_1]$  \\
					&&& $k_{z}[(\lambda_{13}+\lambda_{31})\sigma_1+(\lambda_{23}+\lambda_{32})\sigma_2]$ \\
					&&& $k_{x}k_{y}k_{z}[(\lambda_{13}+\lambda_{31})\sigma_2+(\lambda_{23}+\lambda_{32})\sigma_1]$\\
					&&& $k_{z}(k_{x}^{2}-k_{y}^{2})[(\lambda_{13}+\lambda_{31})\sigma_1-(\lambda_{23}+\lambda_{32})\sigma_2]$\\  \hline
					\multirow{8}{*}{$B_{1u}$}   & \multirow{8}{*}{\shortstack{$k_{z}(k_{x}^{2}-k_{y}^{2})(\lambda_{12}-\lambda_{21}) $ \\
							$k_{x}(\lambda_{13}-\lambda_{31})+k_{y}(\lambda_{23}-\lambda_{32}) $}}
					& \multirow{8}{*}{\shortstack{$k_{z}(k_{x}^{2}-k_{y}^{2})(\lambda_{11}+\lambda_{22})\sigma_3$ \\
							$k_{z}(k_{x}^{2}-k_{y}^{2})\lambda_{33}\sigma_3$ \\
							$k_{z}(\lambda_{11}-\lambda_{22})\sigma_3$ \\
							$k_{x}k_{y}k_{z}(k_{x}^{2}-k_{y}^{2})(\lambda_{12}+\lambda_{21})\sigma_3$ \\
							$[k_{y}(\lambda_{13}+\lambda_{31})-k_{x}(\lambda_{23}+\lambda_{32})]\sigma_3$}} & $(\lambda_{11}+\lambda_{22})(k_{x}\sigma_1+k_{y}\sigma_2)$ \\
					&&& $\lambda_{33}(k_{x}\sigma_1+k_{y}\sigma_2)$ \\
					&&& $(\lambda_{11}-\lambda_{22})(k_{x}\sigma_1-k_{y}\sigma_2)$\\
					&&& $(\lambda_{12}+\lambda_{21})(k_{y}\sigma_1+k_{x}\sigma_2)$\\
					&&& $k_{x}k_{y}k_{z}[(\lambda_{13}+\lambda_{31})\sigma_1-(\lambda_{23}+\lambda_{32})\sigma_2]$\\
					&&& $k_{z}[(\lambda_{13}+\lambda_{31})\sigma_2+(\lambda_{23}+\lambda_{32})\sigma_1]$\\
					&&& $k_{x}k_{y}k_{z}[(\lambda_{13}+\lambda_{31})\sigma_1+(\lambda_{23}+\lambda_{32})\sigma_2]$\\
					&&& $k_{z}(k_{x}^{2}-k_{y}^{2})[(\lambda_{13}+\lambda_{31})\sigma_2-(\lambda_{23}+\lambda_{32})\sigma_1]$\\ \hline
					\multirow{8}{*}{$B_{2u}$}   & \multirow{8}{*}{\shortstack{$k_{x}k_{y}k_{z}(\lambda_{12}-\lambda_{21}) $ \\
							$k_{y}(\lambda_{13}-\lambda_{31})-k_{x}(\lambda_{23}-\lambda_{32}) $}} & \multirow{8}{*}{\shortstack{$k_{x}k_{y}k_{z}(\lambda_{11}+\lambda_{22})\sigma_3$  \\
							$k_{x}k_{y}k_{z}\lambda_{33}\sigma_3$ \\
							$k_{x}k_{y}k_{z}(k_{x}^{2}-k_{y}^{2})(\lambda_{11}-\lambda_{22})\sigma_3$ \\
							$k_{z}(\lambda_{12}+\lambda_{21})\sigma_3$ \\
							$[k_{x}(\lambda_{13}+\lambda_{31})+k_{y}(\lambda_{23}+\lambda_{32})]\sigma_3$}} & $(\lambda_{11}+\lambda_{22})(k_{y}\sigma_1-k_{x}\sigma_2)$ \\
					&&& $\lambda_{33}(k_{y}\sigma_1-k_{x}\sigma_2)$ \\
					&&& $(\lambda_{11}-\lambda_{22})(k_{y}\sigma_1+k_{x}\sigma_2)$\\
					&&& $(\lambda_{12}+\lambda_{21})(k_{x}\sigma_1-k_{y}\sigma_2)$\\
					&&& $k_{x}k_{y}k_{z}[(\lambda_{13}+\lambda_{31})\sigma_2+(\lambda_{23}+\lambda_{32})\sigma_1]$\\
					&&& $k_{z}[(\lambda_{13}+\lambda_{31})\sigma_1-(\lambda_{23}+\lambda_{32})\sigma_2]$\\
					&&& $k_{x}k_{y}k_{z}[(\lambda_{13}+\lambda_{31})\sigma_2-(\lambda_{23}+\lambda_{32})\sigma_1]$\\
					&&& $k_{z}(k_{x}^{2}-k_{y}^{2})[(\lambda_{13}+\lambda_{31})\sigma_1+(\lambda_{23}+\lambda_{32})\sigma_2]$\\  \hline
					\multirow{20}{*}{$E_{u}$}    & \multirow{20}{*}{\shortstack{$(k_{y},-k_{x})(\lambda_{12}-\lambda_{21}) $  \\
							$k_{x}k_{y}k_{z}(k_{x}^{2}-k_{y}^{2})[\lambda_{13}-\lambda_{31},-(\lambda_{23}-\lambda_{32})] $  \\
							$k_{z}(\lambda_{23}-\lambda_{32},\lambda_{13}-\lambda_{31}) $  \\
							$k_{x}k_{y}k_{z}(\lambda_{13}-\lambda_{31},\lambda_{23}-\lambda_{32}) $  \\
							$k_{z}(k_{x}^{2}-k_{y}^{2})[\lambda_{23}-\lambda_{32},-(\lambda_{13}-\lambda_{31})] $}}   & \multirow{20}{*}{\shortstack{$(k_{y},-k_{x})(\lambda_{11}+\lambda_{22})\sigma_3$ \\
							$(k_{y},-k_{x})\lambda_{33}\sigma_3$ \\
							$(k_{y},k_{x})(\lambda_{11}-\lambda_{22})\sigma_3$ \\
							$(k_{x},-k_{y})(\lambda_{12}+\lambda_{21})\sigma_3$ \\
							$k_{x}k_{y}k_{z}(k_{x}^{2}-k_{y}^{2})(\lambda_{23}+\lambda_{32},\lambda_{13}+\lambda_{31})\sigma_3$ \\
							$k_{z}[\lambda_{13}+\lambda_{31},-(\lambda_{23}+\lambda_{32})]\sigma_3$ \\
							$k_{x}k_{y}k_{z}[\lambda_{23}+\lambda_{32},-(\lambda_{13}+\lambda_{31})]\sigma_3$ \\
							$k_{z}(k_{x}^{2}-k_{y}^{2})(\lambda_{13}+\lambda_{31},\lambda_{23}+\lambda_{32})\sigma_3$}}
						& $k_{x}k_{y}k_{z}(k_{x}^{2}-k_{y}^{2})(\lambda_{11}+\lambda_{22})(\sigma_1,-\sigma_2)$ \\  &&&$k_{z}(\lambda_{11}+\lambda_{22})(\sigma_2,\sigma_1)$\\
						&&&$k_{x}k_{y}k_{z}(\lambda_{11}+\lambda_{22})(\sigma_1,\sigma_2)$ \\  &&&$k_{z}(k_{x}^{2}-k_{y}^{2})(\lambda_{11}+\lambda_{22})(\sigma_2,-\sigma_1)$\\
					&&&$k_{x}k_{y}k_{z}(k_{x}^{2}-k_{y}^{2})\lambda_{33}(\sigma_1,-\sigma_2)$ \\
					&&&$k_{z}\lambda_{33}(\sigma_2,\sigma_1)$\\
					&&&$k_{x}k_{y}k_{z}\lambda_{33}(\sigma_1,\sigma_2)$ \\
					&&&$k_{z}(k_{x}^{2}-k_{y}^{2})\lambda_{33}(\sigma_2,-\sigma_1)$\\
					&&&$k_{x}k_{y}k_{z}(k_{x}^{2}-k_{y}^{2})(\lambda_{11}-\lambda_{22})(\sigma_1,\sigma_2)$ \\  &&&$k_{z}(\lambda_{11}-\lambda_{22})(\sigma_2,-\sigma_1)$\\
					&&&$k_{x}k_{y}k_{z}(\lambda_{11}-\lambda_{22})(\sigma_1,-\sigma_2)$ \\  &&&$k_{z}(k_{x}^{2}-k_{y}^{2})(\lambda_{11}-\lambda_{22})(\sigma_2,\sigma_1)$\\
					&&&$k_{x}k_{y}k_{z}(k_{x}^{2}-k_{y}^{2})(\lambda_{12}+\lambda_{21})(\sigma_2,-\sigma_1)$ \\  &&&$k_{z}(\lambda_{12}+\lambda_{21})(\sigma_1,\sigma_2)$\\
					&&&$k_{x}k_{y}k_{z}(\lambda_{12}+\lambda_{21})(\sigma_2,\sigma_1)$ \\  &&&$k_{z}(k_{x}^{2}-k_{y}^{2})(\lambda_{12}+\lambda_{21})(\sigma_1,-\sigma_2)$\\
					&&&$(k_{x},k_{y})[(\lambda_{13}-\lambda_{31})\sigma_1+(\lambda_{23}-\lambda_{32})\sigma_2]$\\
					&&&$(k_{y},-k_{x})[(\lambda_{13}-\lambda_{31})\sigma_2-(\lambda_{23}-\lambda_{32})\sigma_1]$\\
					&&&$(k_{x},-k_{y})[(\lambda_{13}-\lambda_{31})\sigma_1-(\lambda_{23}-\lambda_{32})\sigma_2]$\\
					&&&$(k_{y},k_{x})[(\lambda_{13}-\lambda_{31})\sigma_2+(\lambda_{23}-\lambda_{32})\sigma_1]$ \\ \hline
			\end{longtable}
\endgroup
%-------------------------------------------------

\section{Gap structure}
Motivated by the recent NMR \cite{NMR_2019,NMR_specific_heat_2020} and neutron \cite{neutron_2020} experiments, we only focus on the spin-singlet pairings belonging to $E_g$ in this work. %Of course, the studies of all other pairings are straightforward.
We solve the quasiparticle gap with the pairing given by Eq.4 of the main text. The normal state single-particle Hamiltonian is taken from Ref.~\cite{Mazin_2006,ARPES_2014}. 
\begin{center}
\begin{figure}[h!]
\includegraphics[width=0.8\textwidth]{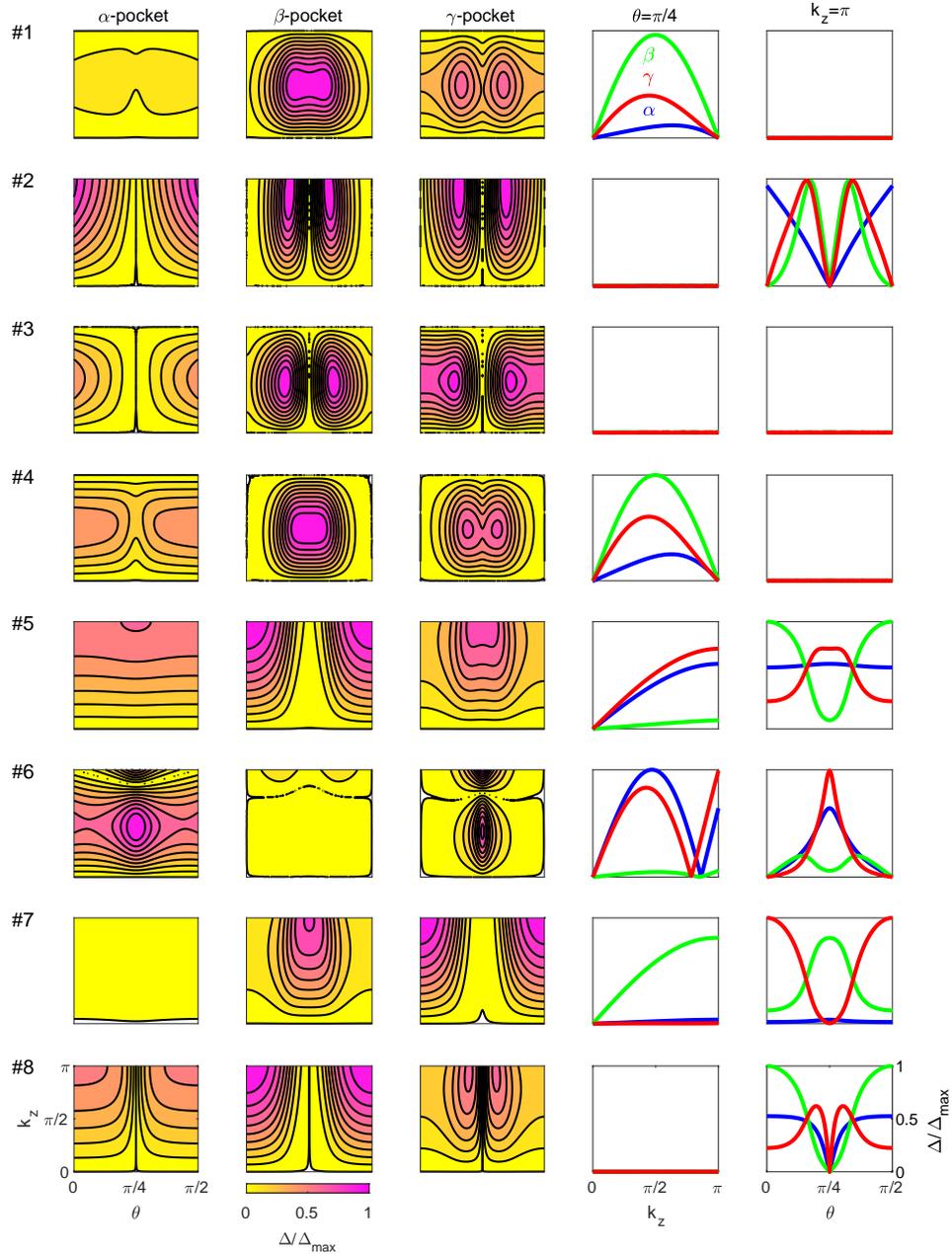}
\caption{Gap structures of each spin-singlet pairing belonging to $E_g$ with TR breaking composition $(d_{xz}+id_{yz})$. The number of each pairing is from the definition in Table I of the main text. In calculations, the value of each pairing is chosen to be $d_i=0.01$mRy (corresponding to about $0.1$meV). For each pairing (line), the first three panels are color plot of the quasiparticle gap on three fermi pockets versus the azimuthal angle $\theta$ and $k_z$. In the fourth and fifth panels, the quasiparticle gaps versus $k_z$ (at $\theta=\pi/4$) and $\theta$ (at $k_z=\pi$) are plotted explicitly.
\label{fig:gap1}
}
\end{figure}
\end{center}

At first, we studied the gap structure of each isolated case with $d_i=0.01$mRy. The results are shown in Fig. ~\ref{fig:gap1}. For each pairing, the quasiparticle gap amplitude contours on three pockets are shown in the first three columns, respectively. In addition, the $k_z$- and $\theta$-dependence are explicitly given in the last two columns. In our plots, $\theta$ is defined as the azimuthal angle relative to $(0,0)$ for $\beta$- and $\gamma$-pockets, while relative to $(\pi,\pi)$ for $\alpha$-pocket. Due to the lattice symmetry, only $0<\theta<\pi/4$ is presented. From these plots, either horizontal or vertical nodal lines can be found. Moreover, due to the inter-orbital pairing, an out-of-plane horizontal nodal line with $k_z\ne0$ is found for No. 6 pairing ($\lambda_{(12)}+\lambda_{(21)}$).

\begin{center}
\begin{figure}[h!]
\includegraphics[width=0.8\textwidth]{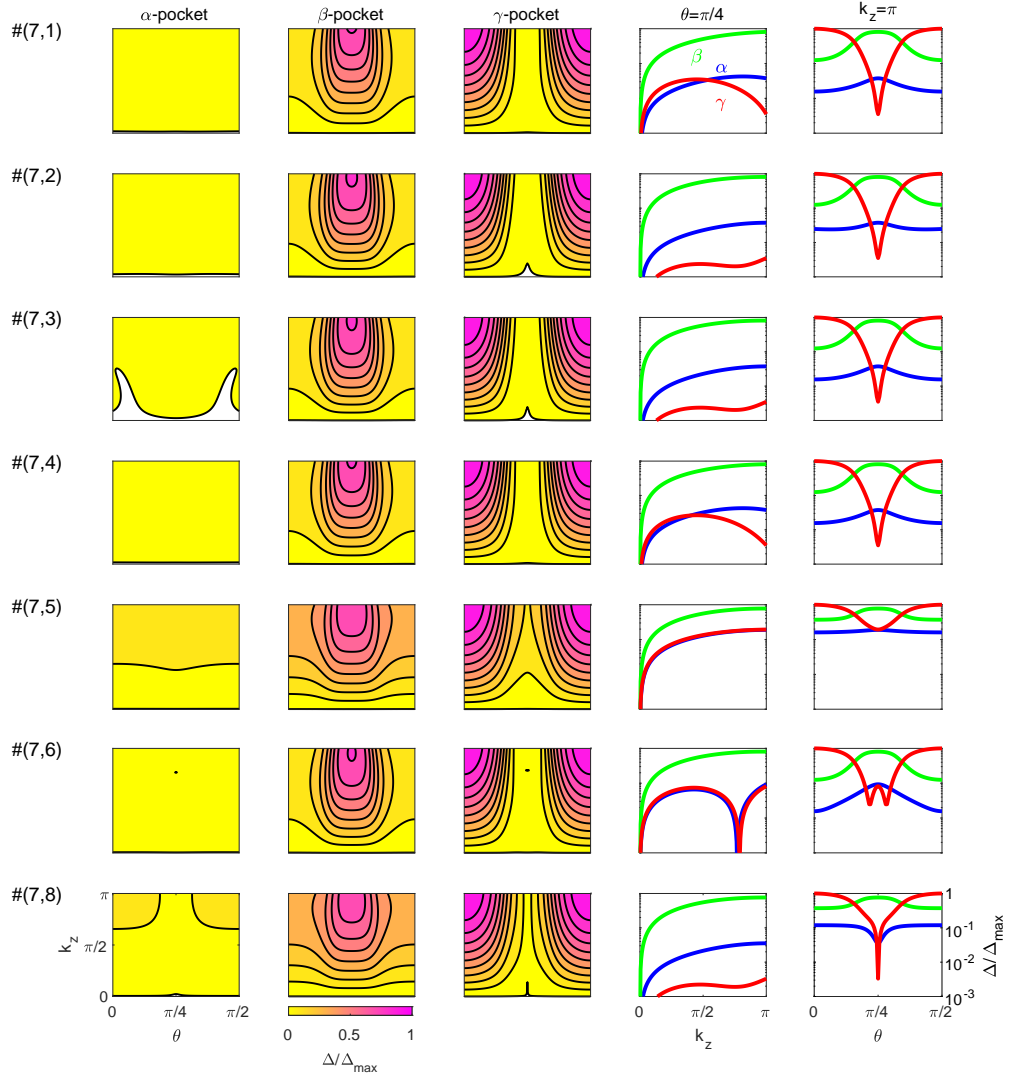}
\caption{Gap structures similar to Fig.~\ref{fig:gap1} but with two pairings coexisting. One pairing is chosen as $d_7=0.01$mRy and the other is $d_i=d_7/2$. Different from Fig.~\ref{fig:gap1}, the last column is plotted with logarithmic scale for clarity.
\label{fig:gap2}
}
\end{figure}
\end{center}

Next, we studied the cases with two pairings coexist in Fig.~\ref{fig:gap2}. We choose $d_7=0.01$mRy and the other component $d_{i\ne7}=d_7/2$ for simplicity. Due to the coexistence of two types of pairings, the vertical nodes are eliminated in general. But for the case of $(d_7,d_8)$, the quasinodes remain along (11)-direction and is compatible with the universal thermal conductivity experiments as discussed in the main text.
Interestingly, in such a multi-orbital pairing with SOC, the gap structure can be very complex. For example, in the case of $(d_7,d_2)$ and $(d_7,d_3)$, we find the original $k_z=0$ horizontal nodal line is extended to a nodal surface called Bogoliubov Fermi surface \cite{Bogoliubov_FS_2017}. While for $(d_7,d_6)$, a nodal point can be found in $\alpha$ and $\gamma$ pockets at $\theta=\pi/4$.

%\end{widetext}

\end{document}